\documentclass[aps,amsart
amsproc,article,elsarticle,emulateapj]{revtex4}
\usepackage{bm}
\usepackage{graphicx,amsmath,latexsym,amssymb}
\usepackage{color}
\usepackage{dcolumn}
\usepackage{pstricks,pst-grad}

\usepackage{pstricks,pst-grad}
\begin{document}
%\baselineskip=24pt
%\vspace{0.5cm}
\title{ Anisotropy effects on the quantum transport of atomic matter waves }
\author{Afifa Yedjour$^{1}$ and Abdel\^{a}ali Boudjem\^{a}a$^{2,3}$}
\affiliation{$^1$Facult\'{e} de Physique, USTO MB, B.P.1505 El M'Naour, 31000, Oran, Alg\'{e}rie\\ 
$^2$Department of Physics, Faculty of Exact Sciences and Informatics, Hassiba Benbouali University of Chlef, P.O. Box 78, 02000, Chlef, Algeria.\\
$^3$Laboratory of Mechanics and Energy, Hassiba Benbouali University of Chlef, P.O. Box 78, 02000, Chlef, Algeria.}
\date{\today}
\large
\begin{abstract}

\begin{center}
 \textbf{Abstract}
\end{center}

We discuss effects of anisotropic scattering in transport properties of ultracold atoms in three-dimensional optical potentials.
Within the realm of the first Born approximation, we calculate  the self energy, the scattering mean free time, the scattering mean free path, and the anisotropy factor.
The behavior of the diffusion constant as a function of the wavenumber is also examined in diffusive and weak localization regimes. 
We show that these quantities are affected by quantum corrections due to the interference caused by disorder.
The dimensionless conductance is also evaluated using the scaling theory of localization.
Our results are compared with previous theoretical and the experimental results.

\end{abstract}

\maketitle
%\textbf{Keywords}: Disorder Amplitude,Mobility Edge, Optical Speckle, Self-Energy.

\section{Introduction}

It has long been proven that a disorder is a crucial ingredient to understand transport properties of waves \cite{1,2}. 
Cold atoms in a disordered  medium have attracted a considerable attention due to their ability of controlling disorder  and interactions \cite{3,4,5,6,7,8,9}. 
One of the most famous phenomena occurring in such disordered systems is the Anderson localization which is characterized 
by a phase transition between a localized regime and a diffusive  regime \cite{2}. 
The localization originates essentially from the interference of multiple-scattering path in the disordered system \cite{11,12}.
Such an interference effect {\it known as weak localization}  leads to increase the  probability to return back to the origin and hence slows down the transport
giving rise to reduce the diffusion constant and the conductivity compared to the Boltzmann or Drude contribution. 
On the other hand, when the localization is strong enough, the wave  remains localized around its initial position guiding to
a total elimination of the transport, which means that the diffusion is stopped and the conductivity is strictly zero. 

Recently, the Anderson transition has been extensively studied both theoretically  \cite{6,113,133,13,14,15,16,17}  and experimentally \cite{18,19,20,21,22}. 
It has been found that the transport of atomic matter waves in three-dimensional (3D) media is significantly
affected by three parameters: the disorder amplitude, the initial wave energy and the correlation energy induced by disorder.
Quite recently, the dynamic evolution of expanding Bose-Einstein condensates (BECs) in 3D random potential has been deeply examined in space and time \cite{23}.

Furthermore, coherent transport and Anderson localization in disordered media can be strongly affected by anisotropy effects (see e.g\cite{Bish,Pine,Nick,Kao,Wier,John,Gur,Wolf,Bart,Stark,24,233,234}).
Recent works \cite{133,13,14,15,233} reveal that the transport anisotropy is reduced near by the mobility edge. 
The anisotropy of the coherent diffusion can be controled  by the properties of the disorder-averaged effective scattering medium.

In this paper, we report on the systematic study of the weak localization effects in the anisotropic transport properties of matter wave in 3D  optical potentials employing
the first Born approximation (FBA).  This latter has been recently proven effective in analyzing the elastic scattering time of matter waves \cite{22}, 
the dynamic evolution of expanding BECs \cite{23}, diffusion and localization in optical disordered potentials (see e.g.\cite{13,24}), 
and good agreement is obtained with experimental and numerical data \cite{22}.
We calculate analytically and numerically the self-energy, the scattering mean free time, the scattering mean free path, and the anisotropy factor.
Our results demonstrate that at low momenta, the scattering mean free time is constant indicating that the scattering is isotropic (i.e in the Boltzmann classical regime).
Whereas at high momenta, it increases linearly with momentum and then diverges pointing out that the scattering reaches the anisotropic regime in agreement with the results of Ref.\cite{14}.
The anisotropy factor increases with  momentum and becomes more pronounced  at large momenta 
revealing  that the scattering is  anisotropic  for all momenta due to the long-range correlations between waves. This is in concordance with the results of \cite{13}.
Moreover, the behavior of the scattering mean free path  emphasizes that the anisotropic effects exists for various momenta at lowest energy $(\varepsilon = 0)$. 
A  crossover from a diffusive regime to a weak localization occurs when the energy becomes comparable with the disorder correlation energy 
due to the long-range correlations between waves.
Correction to the diffusion constant relative to the Boltzmann diffusion constant is also determined in different regimes. 
In weak  localization regime,  we find that the diffusion constant strongly depends on the energy of the system. 
For instance, in the limit of high energy the diffusion constant is reduced. This important manifestation could be explained by the fact that the wave undergoes multiple scattering events 
which evoke a reduced diffusion constant. In the opposite situation where $\varepsilon < 0$, the diffusion constant becomes close to the Boltzmann diffusion constant. 
Additionally, we perform a numerical analysis based on the FBA of the scaling function $\beta$ which is related to the dimensionless conductance 
in order to understand the universality of the flow of bosonic states with the scale transformation.

The remainder of the paper is organized as follows. 
In section \ref{Mod} we calculate the quantities of interest employing the FBA and introduce a brief overview for the speckle potential. 
We write down also the basic equations governing the diffusion of an expanding BEC. 
Section \ref{Num} is devoted to the numerical simulation. We compute the self-energy  and plot its imaginary part. 
We evaluate also the scattering mean time, the anisotropy factor and the diffusion constant.  Our results are compared with previous theoretical and experimental findings. 
In section \ref{DC} we discuss the dimensionless conductance and the $\beta$-function in the diffusive regime. 
Finally, we conclude the paper with a summary in section \ref{Conc}.

\section{Model} \label{Mod}

We consider a weakly interacting BEC of $N\gg1$ atoms of mass $m$ expanding in a 3D random optical potential $U(\textbf{R})$.
The wavefunction $\psi\left( {\bf R},t\right)$ of such a system obeys  the time-dependent Gross-Pitaevskii equation:
\begin{equation}\label{GPE}
i\hbar \frac{\partial \psi}{\partial t}=\left[ -\frac{\hbar^{2}}{2m}\nabla^{2}+ V+U+ g |\psi|^{2}\right]\psi,
\end{equation}
where $g$ is the strength of repulsive interactions between atoms, $V({\bf R})$ is the external trapping potential with trap frequency $\omega_0$,
and  $|\psi({\bf r}, t)|^{2}=n({\bf r}, t)$ is the BEC density in a random potential. 
At time $t > \omega_0^{-1}$ when the disorder is switched on,  the condensate becomes very dilue and thus,  the interactions  can be  neglected ($g\rightarrow 0$) \cite{6}.
In such a case Eq.(\ref{GPE}) reduces to the standard Schr\"odinger equation.\\
We use the FBA in which the complex self-energy $\Sigma(\varepsilon,k)$  of the atom is given by \cite{113,25,26}:
\begin{equation}\label{equa1}
 \Sigma(\varepsilon,k) = \sum_{\mathbf{k}'}\ C(k ,\textbf{k}') [\varepsilon-\varepsilon_{\textbf{k}'}-\Sigma(\ \varepsilon, \textbf{k}')]^{-1},
\end{equation}
where $\sum_\mathbf{k'} \equiv \int d^3k'/(2\pi)^3$, $\varepsilon_{k'} = \hbar^{2} k'^{2}/2m$, and $C (k ,\textbf{k}')$ represents the vertex associated with incoherent (Boltzmann) transport  defined in Fourier space as \cite{13,14}:
\begin{equation}\label{equa2}
C(\textbf{q}) = \int \frac{d \bf R}{(2\pi)^3} \ C({\bf R})\ e^{i \textbf{q} \cdot \textbf{R}},
\end{equation}
where ${\bf q} ={\bf  k'} - {\bf k}$.
The two-point correlation function $C(\textbf{R})$ for a 3D laser speckle potential is given by $C(\textbf{R})=\langle U(\textbf{0}) U(\textbf{R}) \rangle = U_{0}^{2} \mathrm{sinc}^2 (R/\xi)$. 
It is characterized by a typical width  $\xi$ called the correlation length and a disorder amplitude $U_{0}$.

Inserting Eq.~(\ref{equa2}) into (\ref{equa1}), the self-energy can be rewritten as:
\begin{equation}\label{equa3}
\Sigma(\varepsilon, k)=\frac{2\ U_{0}^{2} \xi^{2}}{\pi}  \int^{\infty}_{0} d k' k'^{2} \dfrac{I(k, \textbf{k}')}{\varepsilon- \varepsilon_{\textbf{k}'} - \ \Sigma(\ \varepsilon, \textbf{k}')},
\end{equation}
where 
\begin{equation}\label{equa4}
 I(\ k,\ \textbf{k}') = \int _{0}^{\infty} d R \sin^{2}\left(R/\xi\right) \text{sinc} (k R)\, \text{sinc} (k' R).
 \end{equation}
 Here $I(\ k,\ \textbf{k}')$ gives the main contribution to the correlation function, it vanishes for $ k' > k + 2/\xi $ and for $k' < k - 2/\xi$.
 Keeping in mind that the FBA implies that $k$ is a positive quantity, one has
\begin{equation}\label{equa5}
\text{Im} \Sigma^{0}(k,\varepsilon)= \frac{2\ U_{0}^{2} \xi^{2}}{\pi} \int^{k + 2/\xi}_{0}\ d k' k'^2 \dfrac{ I(k,\textbf{k}')}{\varepsilon- \varepsilon_{\textbf{k}'} + i\ 0}.
\end{equation}
The self-energy $\Sigma(\varepsilon, k)$ is a complex function. Its real part exhibits a logarithmic divergence near $\varepsilon =  \frac{\hbar^{2}}{2m} \big(k + \frac{2}{\xi}\big)^2$.
For $\varepsilon <  \frac{\hbar^{2}}{2m} \big(k + \frac{2}{\xi}\big)^2$, we obtain from Eq.~(\ref{equa5}) the imaginary part of the self-energy:
\begin{equation}
\text{Im} \Sigma^{0}(k,\varepsilon) = - U_{0}^{2} \xi^{2} \int^{k +2/\xi}_{0}\ dk' k'^{2}\ I(k,\textbf{k}') \delta(\varepsilon\ - \ \varepsilon_{\textbf{k}'}).
\end{equation}
At small momenta $k \ll 2/\xi$, one has $\text{Im} \Sigma^{0}(k,\varepsilon) = - \pi U_{0}^{2}/(4\varepsilon_{\xi})$, 
where $\varepsilon_{\xi}=\hbar^2/(2\xi^2m)$ accounts for the quantum correlation energy.
We see that  $\text{Im}\Sigma^{0}(k,\varepsilon)$ is constant, and it depends only on the characteristics of the random potential.
In the limit $k=0$,  the imaginary part of the self-energy can be approximated by a Heaviside step function as:
$\text{Im} \Sigma^{0}(0,\varepsilon) = -(\pi U_{0}^{2}/4\varepsilon_{\xi}) \Theta (\varepsilon-\varepsilon_{\xi})$ \cite{22}, which indicates that some atoms can diffuse
from the optical disordered potential.

In the frame of the FBA which requires the inequality $\varepsilon_{k} \gg U_{0}^{2}/\varepsilon_{\xi}$, the scattering mean-free time is defined as \cite{13,27}:
\begin{equation}\label{equa6}
 \tau_{s}(\varepsilon,k) = \frac{\hbar}{2 \text{Im} \Sigma^{0}(\varepsilon,k)}.
\end{equation}
For an isotropic diffusion, $ \tau_{s}(\varepsilon,k) = 2\hbar\,\varepsilon_{\xi}/ (\pi U_{0}^{2})$.
The scattering mean free path $\ell_{s}(\varepsilon,k)$ and the scattering mean free time are connected with each other via:
$\ell_{s}(\varepsilon,k)= v \tau_{s}(\varepsilon,k)$ where $v= \hbar\ k/m$.
%Inserting into the mean free path $\ell_{s}(\varepsilon,k)$,one obtains $\ell_{s}(\varepsilon,k) = (\frac{\varepsilon}{2m})^{1/2} \frac{2\hbar\varepsilon_{\xi}}{\pi U_{0}^{2}}$.

Then, in the diffusive regime and for weak disorder, the diffusion constant is given by \cite{6,27}:
\begin{equation}\label{equa7}
 D_B =\frac{v\ell_B}{3},
\end{equation}
In the weak-scattering limit $\varepsilon_{k}\backsimeq U_{0}^{2}/\varepsilon_{\xi}$, the diffusion constant is reduced due to the dephasing processes. 
In such a situation, the transport (Boltzmann) mean-free path $\ell_{B}$ is the average distance required to completely erase the memory 
of the initial direction of propagation. For the above 3D disorder speckle, the transport mean-free path is given by 
$ \ell_{B}= (2\ell_{s}/3)\ [(k\xi)^{2}\Theta(k\xi\ - \ 1)  +  \Theta(1\ - \ k\xi)]$ \cite{13}.
 It is related to $\ell_s$ through \cite{13}:
\begin{equation}\label{equa8}
 \frac{\ell_{s}}{\ell_{B}}  = 1\ - \ \langle \cos\theta \rangle,
\end{equation}
where the anisotropy factor $\langle \cos\theta \rangle$ is given by \cite{13}
\begin{equation}\label{equa9}
 \langle \cos\theta\rangle  = \int d \Omega \cos \theta \ {\cal C} (k\xi ,\theta),
\end{equation}
where $\cos \theta$ is averaged over the phase function $ {\cal C} {(k\xi ,\theta)}= \sin \theta\, C(k\xi ,\theta)/  \big[\int d \Omega \,C (k\xi ,\theta)\big]$ with  $C (k\xi ,\theta)$ 
being the angular correlation function versus the scattering angle $\theta$ between $k$ and $k'$ at fixed on-shell momenta $k=k'$, and $d \Omega= 2\pi \sin \theta d\theta$.
For  $\langle \cos\theta\rangle=0$, the scattering is fully isotropic and hence,  $\ell_{B} = \ell_{s}$ \cite{13}. 
Isotropic scattering exists only for $\delta$-correlated disorder (uncorrelated disorder).
Contrary, when $\langle \cos\theta\rangle \neq 0$, the diffusion is anisotropic and  the corrections due to weak localization are pronounced. 
In this case, a large number of scattering events is necessary to deviate the particle completely, thus $\ell_{B} \gg \ell_{s}$ \cite{13}.

The diffusion constant $D_{\varepsilon}$ must be written as \cite{24,25,26} :
\begin{equation}\label{equa10}
 \frac{1}{D_{\varepsilon}} -  \frac{1}{D_{B}} =  \frac{2\hbar}{m k D_{B}}\int_{0}^{\infty} \frac{q\ dq}{q^{2} - i0},
\end{equation}
In 3D, the integral over momentum $q$ in Eq.~(\ref{equa10}) requires only a cutoff at large distances (see e.g. \cite{13,28}) yielding:
\begin{equation}\label{equa11}
 D_{\varepsilon}= D_{B} \left [1- \frac{3}{\pi(k\ell_B)^{2}} \right].
\end{equation}
This equation shows that  the weak localization reduces the diffusion constant due to the corrections induced by the fluctuations caused by the random potential (weak localization effects).
Such interference corrections are small in the weak disorder regime $k \ell_s \gg1$ \cite{13}.

\begin{figure}
\begin{center}
\includegraphics [angle=360,width=0.7\columnwidth]{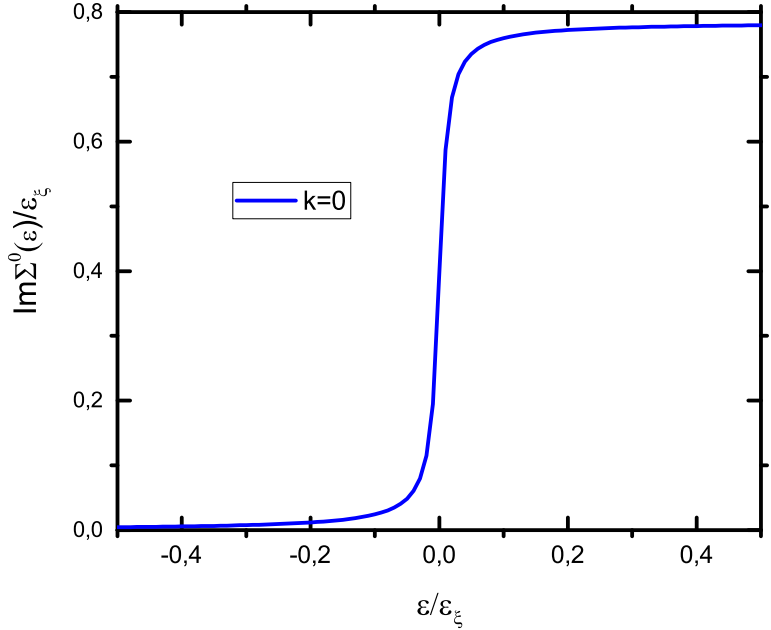}
\caption{Imaginary part of the self-energy $\text{Im} \Sigma^{0}$ for $ U/\varepsilon_{\xi}=1$.}
\label{1.eps}
\end{center}
\end{figure}

\section{Numerical results} \label{Num}

We solve numerically  Eq.~(\ref{equa5}) at $k=0$ for different energies. 
Our calculation is based on the FBA which offers more scattering  channels to particles travelling along the diffusion.\\
Figure \ref{1.eps} shows that  $\text{Im} \Sigma(\varepsilon)$ presents a step function behavior with $\varepsilon$ and strongly depends on the energy
in agreement with the above analytical results. 

\begin{figure}
\includegraphics [angle=360,width=0.7\columnwidth]{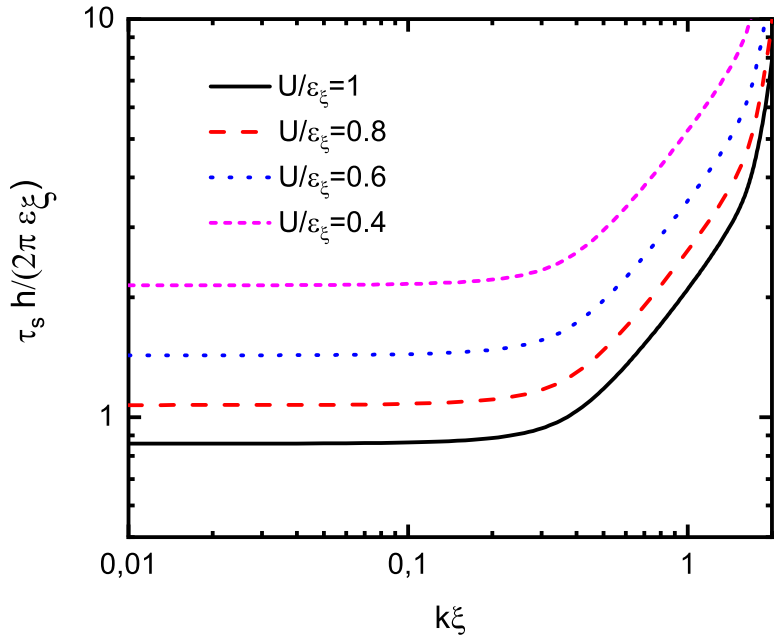}
\caption{The mean free time $\tau_{s}$   normalized by $h/(2\pi\varepsilon_{\xi})$ as a function to momentum for various disorder amplitude on a logarithmic scale.}
\label{2.eps}
\end{figure}

The scattering mean free time is obtained from Eq.(\ref{equa6}). 
Figure \ref{2.eps} depicts that the curves have the same behavior for various disorder
amplitudes and diverge at large momentum in similar fashion with the numerical and experimental findings of  Ref.\cite{22}. 
Within Boltzmann approximation for $k\xi<1$, the contribution of the scatterers in the direction of incident propagation vanishes by the ensemble average over 
all realizations of the random potential giving  rise to an isotropic $\tau_{s}$. 
Our results show that for small momentum, the scattering mean free time converges to a constant values $\tau_{s}=0.96\hbar/\varepsilon_{\xi}$ until $k\xi=0.2$ in agreement with the
theoretical results of Ref.\cite{14}.
For $k\xi \gtrsim 0.2$, the scatterers start to deviate for all values of disorder amplitude.
Another important remark is that as the disorder amplitude increases, $\tau_{s}$ increases linearly with  $k\xi$ in the high-momenta limit.
It is also shorter for large amplitudes in the whole range of $k\xi$.

\begin{figure}
\includegraphics [angle=360,width=0.7\columnwidth]{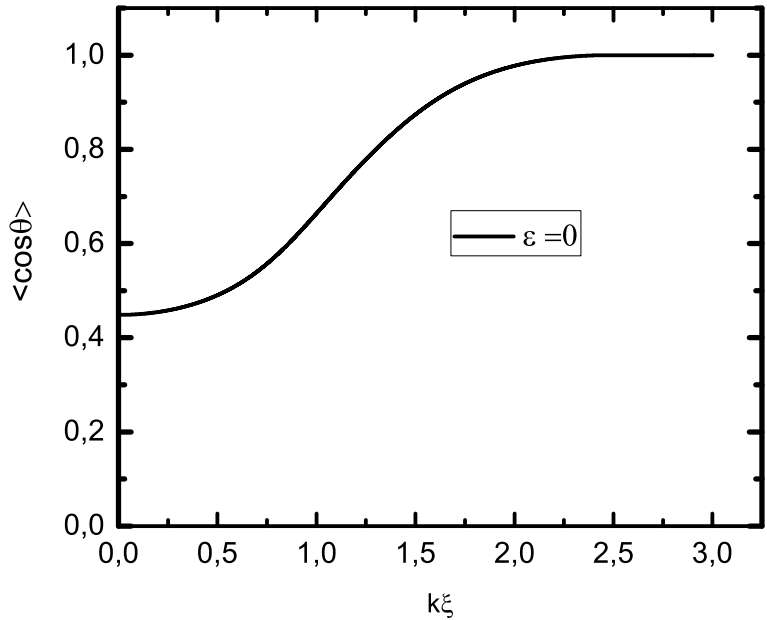}
\caption{Anisotropic factor as function of momentum. At lowest momentum $\langle\cos\theta \rangle =0.45$ emphasizing the anisotropic effects.}
\label{3.eps}
\end{figure}

Actually according to Boltzmann approximation for $k\xi \gg 1$,  the scattering is strongly deviated forward in a scattering cone of an angle $\theta$. 
The deviation which regards only the faster atoms may modify the trajectory of  the waves during the diffusion.
The scattering angle $\theta$ which is defined as the angle between the initial wave vector $k$ and the final wave vector $k'$, is averaged over the phase function $\cos\theta$.
For $k\xi \gg 1$, one has $\theta \ll 1$ yielding $\cos\theta =1-\theta^{2}/2$.
Therefore, equation (\ref{equa8}) gives $\tau_{B} = 2\tau_{s}/\langle\theta^{2} \rangle$ 
marking that when forward scattering is enhanced, the Boltzmann time largely exceeds the Drude time.
For $k\xi=1$, $\tau_{s}=1.3\hbar/\varepsilon_{\xi}$ which is bigger than the initial momentum ($\tau_{s}=0.96\hbar/\varepsilon_{\xi}$) caused by the dephasing process.
The deviation can be attributted to the velocity of the atom which causes fluctuations in its trajectory showing that $\tau_s$ exhibits an anisotropy behavior.

\begin{figure}
\includegraphics[angle=360,width=0.7\columnwidth]{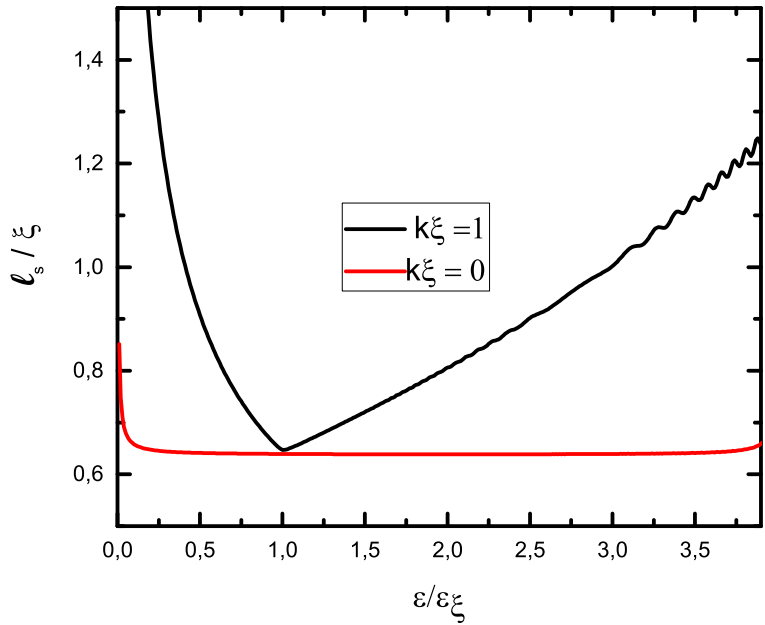}
\caption{The scattering mean free path $\ell_s$  as a function of the $ \varepsilon / \varepsilon_{\xi}$ for $U/\varepsilon_{\xi}$ =1.}
\label{4.eps}
\end{figure}

The anisotropy factor $\langle \cos\theta \rangle$ as a function of the wavenumber $k\xi$ is captured  in figure \ref{3.eps}. 
We see that for very small wave $k\xi\sim0$, $\cos\theta=0.45$ confirming that the anisotropic effects persist even at small momenta.
The numerical calculation gives $\tau_{s} = 0.55 \tau_{B}$ at $k\sim0$ indicating that some atoms are sensitive to dephasing processes and therefore correspond to diffusive propagation.
The anisotropy factor increases with  $k\xi$ and becomes noticeable ($\langle \cos\theta \rangle \sim 1$) at large momenta. 
We infer that the scattering is  anisotropic at lowest energy $(\varepsilon=0)$ for all momenta due to the fact that the disorder is not pointlike. 
It is thus related to the finite disorder correlation length.
These results are comparable with those found in the literature \cite{23}, where $\langle \cos\theta \rangle = 1/3$ for all $k\xi \ll 1$.

\begin{figure}
\centering{
\includegraphics [angle=360,width=0.45\columnwidth]{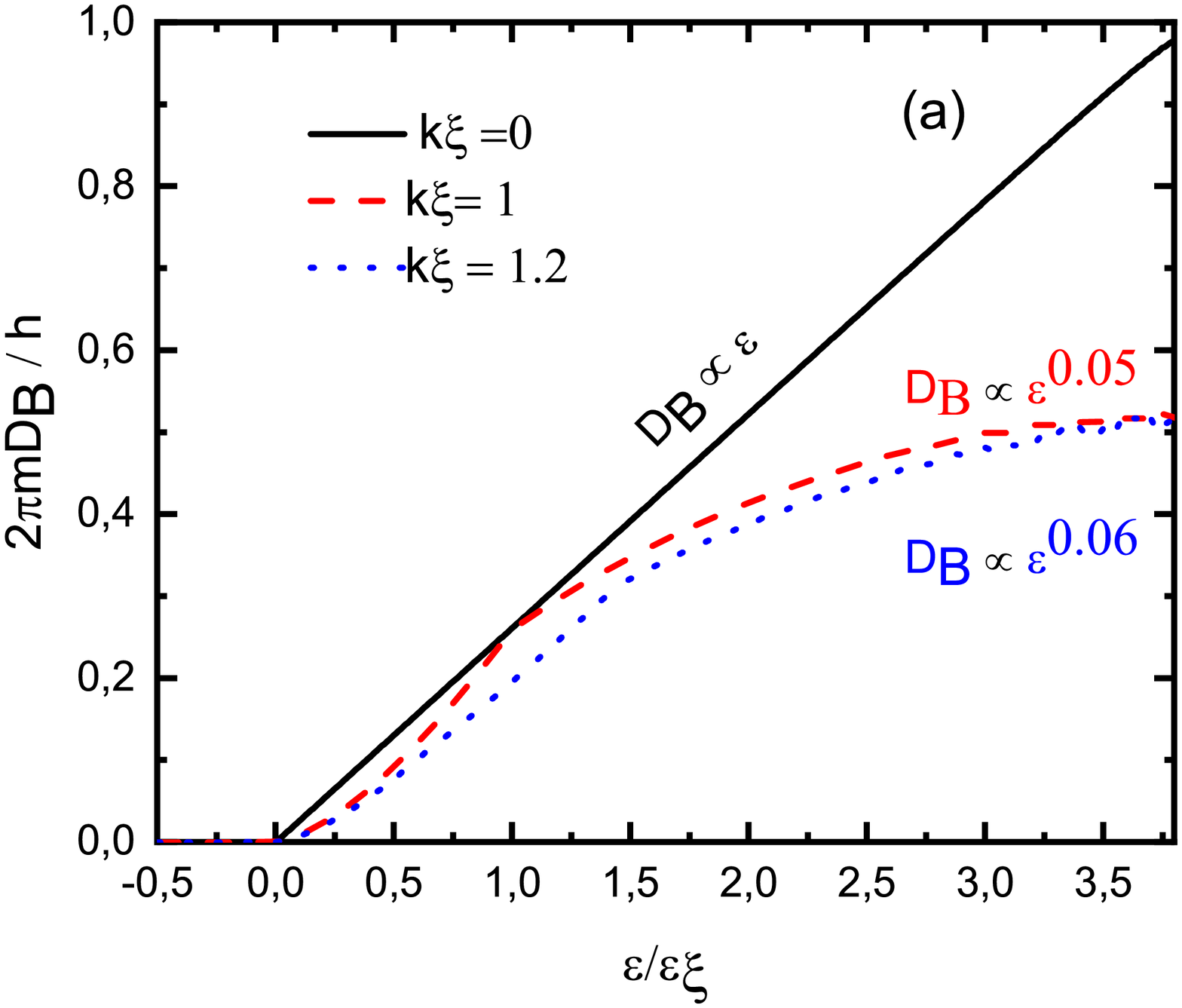}
\includegraphics [angle=360,width=0.45\columnwidth]{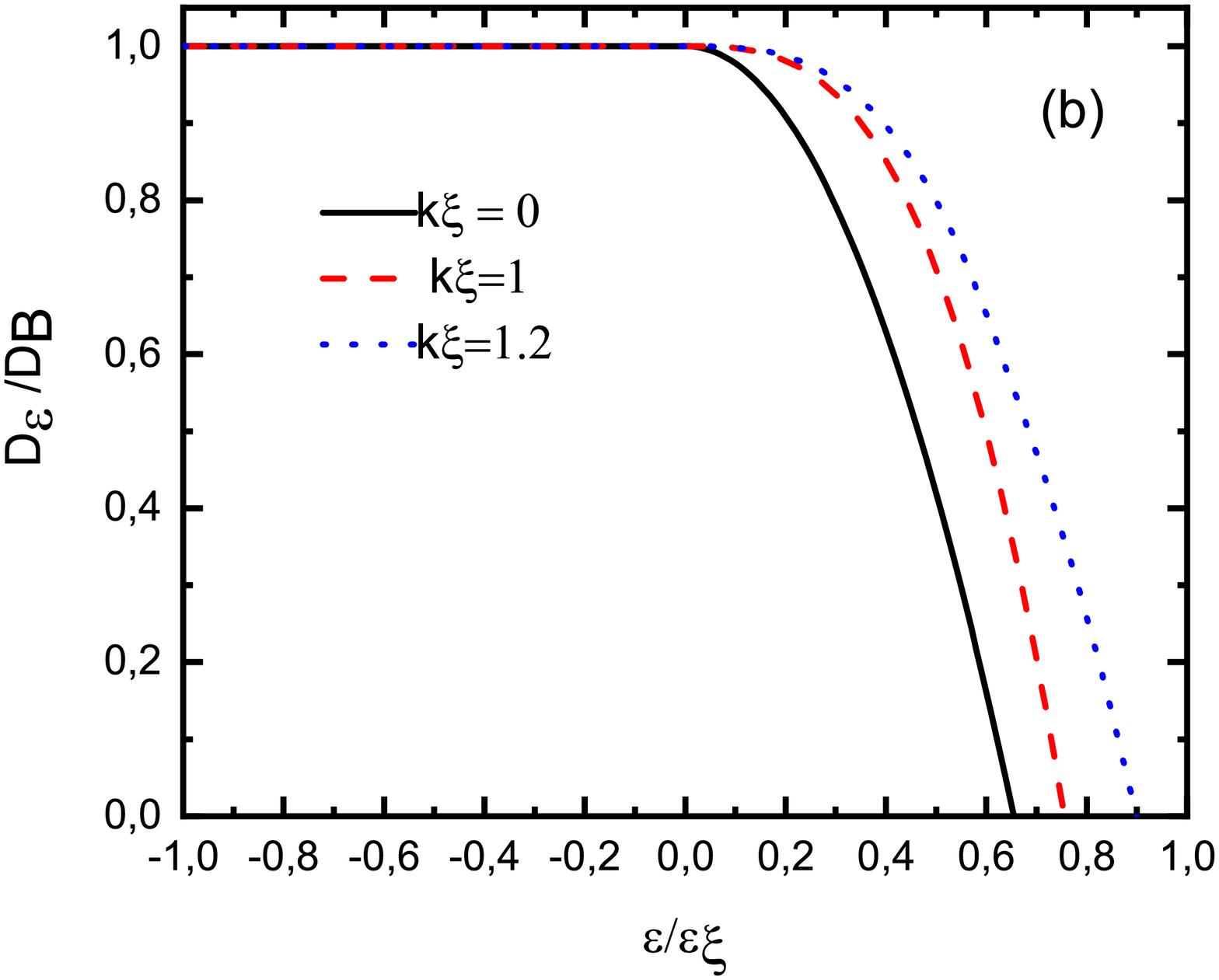}}
\caption{ (a) Boltzmann diffusion constant normalized by $(2\pi m)/h$ as a  function of the energy  for three different values of momenta.
At $\varepsilon=0.1\varepsilon_{\xi}$ deviations in $k\xi =1$ and $k\xi =1.2$ are due  to week localzation effects.
(b)  Weak localization correction relative to the Boltzmann diffusion constant versus $\varepsilon / \varepsilon_{\sigma}$ at $\varepsilon=0$ for 
for three different values of $k\xi$.}
\label{5.eps}
\end{figure}

Now let us analyze the scattering mean free path $\ell_{s}$ in terms of the pertinent parameter $\varepsilon/\varepsilon_{\xi}$.
Figure \ref{4.eps} shows that $\ell_{s}$ first decreases at  lower energies ($\varepsilon < \varepsilon_{\xi}$) until it reaches $\ell_{s} = 0.66\xi$ at 
$\varepsilon = \varepsilon_{\xi}$ signaling the presence of the residual anisotropy. 
In the opposite, at higher energies ($\varepsilon > \varepsilon_{\xi}$), $\ell_{s}$ increases  displaying a strong anisotropy for fast atoms as predicted in Refs.\cite{13,14}.
This nonmonotonic behavior would indicate that the long-range correlations in the optical potential may substantially affect the scattering anisotropy.\\
The situation is quite different at the initial state ($k\xi\simeq 0$), where the scattering mean free path $\ell_{s}$ decreases 
rapidly at $\varepsilon\sim 0$ towards the same value $\ell_{s} = 0.66\xi$ and it remains constant regardless of the particle energy.
Our analysis based on the anisotropy factor shows that $\ell_{s} = 0.55\ell_{B}$ which is close to the results of \cite{13} ($\ell_{s} = 2\ell_{B}/3$).
We conclude that the anisotropic character in 3D case is always present during the diffusion process.

%\begin{figure}
%\includegraphics [angle=360,width=0.7\columnwidth]{6.eps}
%\caption{ Weak localization correction relative to the Boltzmann diffusion constant versus $\varepsilon / \varepsilon_{\sigma}$ at $\varepsilon=0$ for 
%for three different values of $k\xi$.}
%\label{6.eps}
%\end{figure}

The Boltzmann diffusion constant in the diffusive regime can be computed through Eq.~(\ref{equa7}). 
Its behavior as a  function of the energy  for three different momenta is displayed in  figure \ref{5.eps} (a).
It is clearly visible that at very low momenta $k\simeq 0$,  the diffusion constant is a linear function of the energy, $D_B\propto \varepsilon$,
for $0 <\varepsilon/\varepsilon_{\xi}<4$ indicating the persistence of the anisotropy of $D_B$ even at very low momenta. 
This emphasizes the absence of a 3D white-noise limit (i.e. nonexistence of isotropic diffusion)  in good agreement with the results of \cite{14}.
However, for $k\xi = 1$ and $k\xi =1.2$,  $D_B$ increases as a power-law $\propto \varepsilon^{\alpha}$ ($\alpha=0.05$ for $k\xi = 1$ and $\alpha=0.06$ for $k\xi =1.2$)
in the region $\varepsilon/\varepsilon_{\xi} \lesssim 3.5$. 
For $\varepsilon \gtrsim 3.5\varepsilon_{\xi}$, the diffusion constant begins to saturate at $2\pi m D_B/h \approx 0.46$.
The two curves for $k\xi = 1$ and $k\xi = 1.2$ match at $\varepsilon/\varepsilon_{\xi} \lesssim 3.8$.
This can be understood by the fact that the anisotropy factor has a non-negligeable effect for $k\xi \neq 0$ as already indicated in figure \ref{3.eps} giving rise to 
a significant anisotropic diffusion constant.

In order to examine the role of weak localization effects around the region $\varepsilon=0$ in the diffusion constant, we solve numerically Eq.~(\ref{equa11}). 
The results are shown in figure \ref{5.eps} (b).
We observe that the diffusion constant is compatible with the Boltzmann diffusion constant $(D_{\varepsilon}=D_{B})$ for $\varepsilon/\varepsilon_{\xi} \lesssim 0$.
Then for $\varepsilon/\varepsilon_{\xi}>0$, the diffusion constant decays as $D_{\varepsilon} < D_{B}$ for all momenta owing to the  weak localization effects. 
We see also that for $k\xi\sim 0$, the wave form multiple scattering paths that are first  affected by the correlation  leading to the diminution of the diffusion constant. 
This latter falls quickly to zero for the three values of momentum. For instance, at $\varepsilon/\varepsilon_{\xi}=0.5$, one has  $D_{\varepsilon} \simeq 0.4 D_{B}$ for $k\xi=0$,
while this value rises to $D_{\varepsilon} \simeq 0.8 D_{B}$ for $k\xi=1$ and to $D_{\varepsilon} \simeq 0.85 D_{B}$ for $k\xi=1.2$.

\begin{figure}
\includegraphics [angle=360,width=0.7\columnwidth]{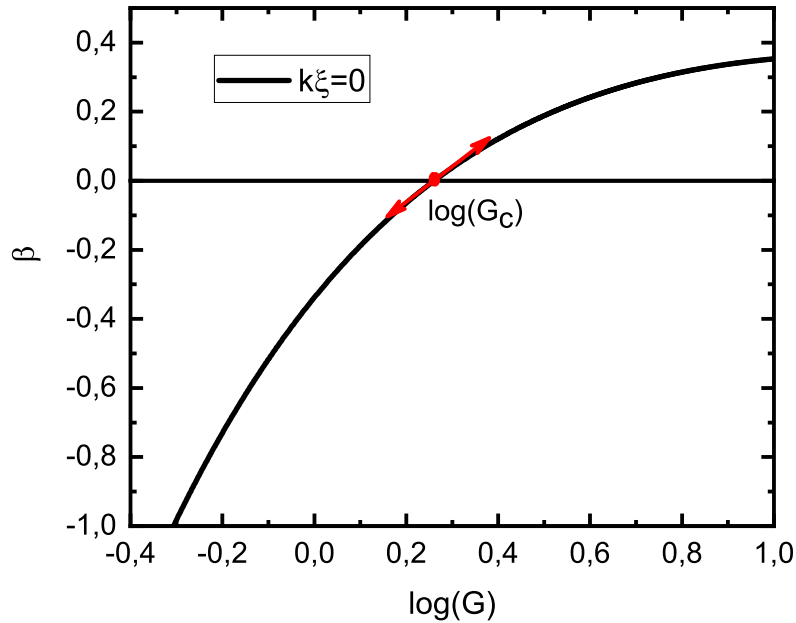}
\caption{ Conductance scaling $\beta$-function.
The crossing with the horizontal axis indicates the transition point between the diffusive and the localized regimes.}
\label{GL}
\end{figure}

\section{Dimensionless conductance and $\beta$-function} \label{DC}

In order to check the universal transport properties, we use the scaling theory which is formulated in terms of a dimensionless conductance.
In the diffusive regime, the dimensionless conductance $G$ of a sample of linear size $L$ is defined as: $G= 2 (k \ell_B) (k L)/3$ (see e.g \cite{28}).
Our aim is to look at how the dimensionless conductance evolves with system size, then it is useful to introduce the scaling $\beta$-function \cite{2,28} 
\begin{equation}\label{beta}
\beta=  \frac{d \ln G}{d \ln L},
\end{equation}
which shows the flow of bosonic states. For $\beta=0$,  the conductance $G$ does not vary with $L$. The results of the numerical analysis of equation (\ref{beta}) based on the FBA
for the function $\beta$ are sketched in figure \ref{GL}.
We see that $\beta$ is negative for small $G$ while it is positive for large values of $G$.  
It crosses zero at $G_c \approx 1.2$ which is close to the theoretically predicted value $G_c \approx 1$.
This unstable fix-point can be considered as a critical point marking a possible transition from a diffusive regime to a localized regime.
%The small discrepancy between the two values is most probably due to the interference effects.

\section{Conclusions} \label{Conc}

We investigated the anisotropy effects on the quantum transport of atomic matter waves using the FBA. 
We first calculated  analytically and numerically the imaginary part of self-energy which leads us to examine the scattering mean time of diffusion.
Our results indicate that the scattering time is inversely proportional to the square of the disorder amplitude regardless of the values of the energy in good agreement with recent experiments.
A deviation of $\tau_s$ from isotropy exhibits an angular dependence  (i.e. depends on the anisotropy angle $\theta$).
The anisotropy of the scattering mean free path is revealed in its nonmonotonic character induced by the long-range correlations of the optical potential.
Strictly speaking, the scattering is never isotropic ($l_s \sim 0.45 l_B$) for small momenta $k\xi \sim 0$ even though  the anisotropy decreases ($\langle \cos \theta \rangle \rightarrow 0$).
While for high momenta $k\xi \gg 1$, the anisotropy is singificant results in strong scattering anisotropy.

Moreover, the Boltzmann diffusion constant $D_B$ has been explored numerically. 
At higher energy, we found that  $D_B$  increases linearly in agreement with theoretical findings of Refs.\cite{13,14}.
However, it follows as power law for $k\xi=1$ and $k\xi=2$ and is significantly anisotropic due to the correlations induced by the multiple scattering in the optical disorder.
We then analyzed the effects of weak localization on the deviation of the diffusion constant $D_{\varepsilon}/D_B$.  Our predictions show that for $\varepsilon/\varepsilon_{\xi} \leq 0$, 
$D_{\varepsilon}=D_B$ for all momenta. At higher momenta and for $\varepsilon/\varepsilon_{\xi}> 0$, the weak localization corrections are strong enough to decrease the ratio $D_{\varepsilon}/D_B$.

Furthermore, we found that the monotonic variation of the conductance with $L$ is prominent  confirming that there is a universal quantum transition from 
a diffusive regime to a localized regime.

We believe that our findings may have implications for understanding the properties of coherent transport in 3D anisotropic optical disorders and provide good reference data for future research.

\end{document}